\documentclass[journal=jpcbfk,manuscript=article,layout=twocolumn]{achemso}

\usepackage{amsmath}
\usepackage{rotating}
\usepackage[usenames, dvipsnames]{color}
\usepackage{booktabs}
\usepackage{multirow}
\usepackage{caption}
\usepackage[normalem]{ulem}

\newcommand{\jdc}[1]{#1} 

\newcommand*{\half}{\ensuremath{\tfrac{1}{2}}}
\newcommand*{\Normal}{\mathcal{N}}
\newcommand*{\hamil}{\mathcal{H}}
\newcommand*{\dt}{\Delta t}
\newcommand*{\work}{W}
\newcommand*{\heat}{Q}
\newcommand*{\shadowwork}{\work_{\rm shad}}			
\newcommand*{\protwork}{\work_{\rm prot}}			
\newcommand*{\PE}{\mathcal{U}}
\newcommand*{\KE}{\mathcal{T}}
\newcommand*{\traj}{X} 
\newcommand*{\PP}[2]{P\big[\, #1\, \big|\, #2\,\big]}
\newcommand*{\x}{r}	
\newcommand*{\wiener}{\, \mathsf W} 
\newcommand*{\f}{\frac}
\newcommand*{\quar}{\tfrac{1}{4}}
\newcommand*{\tquar}{\tfrac{3}{4}}
\def\l{\left}
\def\r{\right}

\author{David A.\ Sivak}
\email{david.sivak@ucsf.edu}
\altaffiliation{Current address: Center for Systems and Synthetic Biology, University of California, San Francisco, California 94158, USA}
\affiliation{Physical Biosciences Division, Lawrence Berkeley National Laboratory, Berkeley, California 94720, USA}

\author{John D.\ Chodera}
\affiliation{Computational Biology Program, Memorial Sloan-Kettering Cancer Center, New York, New York 10065, USA}

\author{Gavin E.\ Crooks}
\affiliation{Physical Biosciences Division, Lawrence Berkeley National Laboratory, Berkeley, California 94720, USA}

\title[Time Step Rescaling Recovers Continuous-Time Langevin Dynamical Properties]{Time Step Rescaling Recovers Continuous-Time Dynamical Properties for Discrete-Time Langevin Integration of Nonequilibrium Systems}

\begin{document}

\small

\begin{abstract}
When simulating molecular systems using deterministic equations of motion (\emph{e.g.}, Newtonian dynamics), such equations are generally numerically integrated according to a well-developed set of algorithms that share commonly agreed-upon desirable properties. However, for stochastic equations of motion (\emph{e.g.}, Langevin dynamics), there is still broad disagreement over which integration algorithms are most appropriate. While multiple desiderata have been proposed throughout the literature, consensus on which criteria are important is absent, and no published integration scheme satisfies all desiderata simultaneously.  Additional nontrivial complications stem from simulating systems driven out of equilibrium using existing stochastic integration schemes in conjunction with recently-developed nonequilibrium fluctuation theorems. Here, we examine a family of discrete time integration schemes for Langevin dynamics, assessing how each member satisfies a variety of desiderata that have been enumerated in prior efforts to construct suitable Langevin integrators. We show that the incorporation of a novel time step rescaling in the deterministic updates of position and velocity can correct a number of dynamical defects in these integrators. Finally, we identify a particular splitting \jdc{(related to the velocity Verlet discretization)} that has essentially universally appropriate properties for the simulation of Langevin dynamics for molecular systems in equilibrium, nonequilibrium, and path sampling contexts. 

\end{abstract}

\section{Introduction}
Simulating the dynamics of molecular systems on a digital computer requires that the equations of motion be discretized. The resulting discrete-time integration algorithm that governs the updates of particle positions and velocities will necessarily have properties that differ from the continuous equations of motion on which it was based. To construct such an algorithm, one must decide which properties of the original dynamics should be preserved. Even then, a multitude of integration schemes may satisfy these properties and still recover the continuous stochastic equations of motion in the limit of an infinitesimally small time step.

For integrating the deterministic classical equations of motion prescribed by Newtonian dynamics, explicit symplectic integration schemes such as velocity Verlet are now widely regarded as being optimal for condensed matter systems for a number of reasons: they are reversible, simple to implement, preserve phase space volume, require minimal force evaluations, and are generally stable over long integration times~\cite{Swope1982,AllenTildesley,Frenkel2002,Leach,Schlick00}. 

For stochastic equations of motion---in which the influence of some system components (often the solvent) are not represented explicitly, but instead by random collisions with fictitious particles---no such generally-adopted integrator yet exists. In particular, the dynamics produced by existing algorithms differ (in a time step dependent manner) in important respects from the dynamics of the continuous equations of motion. Nevertheless, significant effort has been devoted to developing such stochastic integrators due to their utility in simulating many systems of interest to the chemical, biophysical, and physical sciences.

Driven nonequilibrium systems present their own set of special challenges. For example, a powerful set of nonequilibrium work fluctuation theorems~\cite{Jarzynski:2011hl} permit the computation of equilibrium properties of systems from their nonequilibrium statistics, but require as input the distribution of work values associated with the ensemble of trajectories. Calculations that use naive analogies of the work for continuous dynamics---failing to take into account details of the discrete integration scheme---possess systematic biases~\cite{Sivak:2013ke}. 

Here, we consider various choices that could be made in constructing discrete-time integration schemes for Langevin dynamics. By examining the various possible Strang splittings of the Langevin Liouville operator, incorporating a novel time step rescaling 
[Eq.~\eqref{tau}], 
and comparing their resulting properties to several desiderata that have been enumerated in the literature over recent decades, we show how it is possible to simultaneously satisfy nearly all these criteria with a single integration scheme 
[Eq.~\eqref{equ:OVRVO}] 
that is generally applicable, simple to implement, computationally efficient, and produces thermodynamically-consistent nonequilibrium statistics.

\section{Theory}

\subsection{The Langevin equation}
A standard framework for the stochastic simulation of molecular systems assumes that the variables of interest evolve according to Langevin dynamics with uncorrelated Gaussian noise~\cite{Langevin1908}, which represents interactions with the surrounding environment through frictional drag and stochastic \jdc{collisions with fictitious bath particles}:
\begin{subequations}
\label{equ:contLang}
\begin{align}
d\x & = v\ dt \\
dv & = \frac{f(t)}{m} \,dt - \gamma v\ dt + \sqrt{ \frac{2\gamma}{\beta m}  }\ d\wiener(t) \ .
\end{align}
\end{subequations}
Here $\x$ and $v$ are time-dependent position and velocity, $m$ is mass, $\beta=1/ k_{\text{B}}T$, $k_\text{B}$ is Boltzmann's constant, $T$ is the temperature of the environment, $\gamma$ is a friction coefficient (with dimensions of inverse time), and $\wiener(t)$ is a standard Wiener process. The force $f(t)$ is due to the (in general  time-dependent) Hamiltonian $\hamil(t)$ on the system with position $\x(t)$, as determined by the derivative of the potential energy, $-\,\partial\hamil(t)/\partial \x$, evaluated at $\x(t)$. For multi-dimensional, multi-particle systems, $\x$, $v$, $f$, and $d\wiener$ are vectors, and $m$ is a diagonal matrix (see Supporting Information, ``Multiple dimensions.'') 

While these equations of motion can be solved exactly for some simple systems, nearly all complex systems of interest require computational techniques in order to generate dynamical trajectories. On digital computers, this requires discretization of time. The selection of an appropriate discrete time integration scheme is made difficult by the fact that \emph{many} discretizations may exist that recover the same continuous stochastic differential equations of motion in the limit of an infinitesimally small time step, but these schemes may possess very different properties for finite time steps.

\subsection{Desiderata}
\label{sec:Desiderata}
Finite time step integrators for molecular systems cannot hope to \emph{exactly} reproduce \jdc{snapshots from the dynamical} trajectories of the continuous equations of motion \jdc{of a real physical system}. Imprecision of experimental measurements ensures that simulated initial conditions necessarily deviate from `true' ones, and Lyapunov instability of even deterministic dynamics ensures the rapid chaotic growth of such deviations, making the exact reproduction of a particular trajectory impossible even were it desirable~\cite{Frenkel2002}. Moreover, artifacts are inevitably introduced when one discretizes continuous equations of motion in a straightforward manner: dynamical motion increasingly diverges from that of continuous equations of motion with increasing friction and/or time step. Instead, the goal is often to reproduce certain statistical (often observable) properties of the system, especially in terms of correlation functions and (possibly time-dependent) ensemble expectations given a set of initial conditions. Thus, a desirable approximation scheme should share certain statistical and dynamical properties of the ensemble of trajectories associated with the exact equations of motion, in lieu of being able to exactly integrate trajectories. 

Pastor, \emph{et al.}~\cite{Pastor:1988dm} proposed that a useful discrete-time integrator should reproduce seven quantities associated with the continuous-time equations of motion: for a free particle (zero-force), the mean-squared displacement (MSD) as a function of time, the mean-squared velocity (MSV), and the velocity autocorrelation function (VAC); for a uniform external force, the terminal velocity; and for a harmonic potential (linear force), the MSD, MSV, and the virial. In Table~\ref{tab:desiderataDefs}, we define these desired dynamical properties and list their analytically computed values for the continuous equations of motion. In Supporting Information, ``Determination of rescaling parameters,'' we describe in detail the calculation of these quantities for our family of integrators. 
\setlength{\tabcolsep}{2pt}
\begin{sidewaystable*}
\centering
\begin{tabular*}{\textwidth}{cccccc}
External Force & Quantity & \multicolumn{3}{c}{Expression} & Continuous-Limit Value \\
\noalign{\smallskip}\hline\noalign{\smallskip}
& mean-squared displacement & $\langle \x^2(n) \rangle$ & or & $\langle \x^2(n+\half)\rangle$ & $\f{2}{\beta m\gamma} n\dt$ \\
zero & mean-squared velocity & $\langle v^2(n) \rangle$ & or &  $\langle v^2(n+\half) \rangle$ & $\f{1}{\beta m}$ \\
& velocity autocorrelation & $\langle v(n) v(n+n') \rangle$ & or & $\langle v(n+\half) v(n+\half+n') \rangle$ & $\f{1}{\beta m} e^{-\gamma n'\dt}$ \\
\noalign{\smallskip}\hline\noalign{\smallskip}
uniform, $f$ & terminal drift & $\f{\big\langle \x\big(n+1\big) - \x\big(n\big) \big\rangle}{\dt}$ & $=$ & $\f{\big\langle \x\big(n+\half\big) - \x\big(n-\half\big) \big\rangle}{\dt}$ & $\f{f}{m\gamma}$ \\
\noalign{\smallskip}\hline\noalign{\smallskip}
& mean-squared displacement & $\langle \x^2(n) \rangle$ & or & $\langle \x^2(n+\half) \rangle$ & $\f{1}{\beta k}$ \\
linear, $-k\x$ & mean-squared velocity & $\langle v^2(n) \rangle$ & or & $\langle v^2(n+\half) \rangle$ & $\f{1}{\beta m}$ \\
& virial & $m\langle v^2(n) \rangle - k\langle \x^2(n) \rangle$ & or & $m\langle v^2(n+\half) \rangle - k\langle \x^2(n+\half) \rangle$ & 0
\end{tabular*}
\caption{Definition of dynamical properties. \textnormal{Angled brackets denote an average over the ensemble of phase-space trajectories produced by a given integration scheme. We consider an integration scheme to preserve a given dynamical property as long as the scheme produces the continuous-limit value of that dynamical property at some point during the full time step, even if it produces different values at other points.}}
\label{tab:desiderataDefs}
\end{sidewaystable*}

One should not expect a discrete algorithm to give meaningful results on timescales less than a single time step. Those users who prefer to treat a discrete algorithm as a black box will be most interested in integrators that satisfy given dynamical properties at integer time steps. Those willing to look further under the hood would need to know at which specific point within a given time step to measure a given dynamical quantity to recover (or most closely approximate) the continuous-limit value. Thus, we examine dynamical properties at both integer and fractional time steps.

There are several other criteria that one may want an integrator to satisfy, not the least \jdc{of which} being ease of implementation and analysis. Additionally, an integrator that is computationally efficient should have an accuracy that scales reasonably with time step length (here, quadratically, the same accuracy order as popular symplectic integration schemes for deterministic dynamics such as velocity Verlet), permitting relatively large time steps; minimize the number of force evaluations (one per time step) so as to minimize computational effort; and easily incorporate constraints (typically reflecting covalent chemical bonds to light elements such as hydrogen) that \jdc{push the integrator stability limit to larger time step}. Path sampling~\cite{stoltz:j-comput-phys:2007:path-sampling} or path reweighting~\cite{minh-chodera:jcp:2009:path-reweighting,chodera:jcp:2011:dynamical-reweighting} strategies often require an integrator that induces an irreducible Markov chain (\emph{i.e.}, it is possible to transition from any phase space point to any other in a single time step through specific choice of the random variables~\cite{Lelievre2010a}) and that for a given trajectory has a readily evaluated path action that governs the probability of that trajectory within the \jdc{dynamical} ensemble. 

Indeed, the task is further complicated when the integrator must produce thermodynamically-consistent nonequilibrium simulations. To facilitate the use of nonequilibrium fluctuation relations~\cite{Jarzynski:2011hl} and estimators derived from them, the integrator must properly split dynamics into stochastic, explicit Hamiltonian-update, and deterministic substeps that distinguish between the heat, work, and shadow work.\bibnote{The \emph{shadow work} denotes the net energy change during deterministic symplectic substeps that, due to their discretization, do not preserve the system Hamiltonian. For a detailed discussion of shadow work and its importance in nonequilibrium fluctuation relations, see Ref.~\citenum{Sivak:2013ke}} For the calculated works to be thermodynamically meaningful, the integrator must also have a form symmetric under time-reversal. 

Pastor, \emph{et al.} demonstrate that no member of their family of overdamped integrators can simultaneously satisfy more than four of their seven desired dynamical properties. Many other underdamped discretizations of the Langevin equation have been proposed~\cite{Brunger1984,Athenes2004,Adjanor2005,Adjanor2006,Vanden-Eijnden2006,Bussi2007,Bussi2009,Izaguirre:2010ve,Bou-Rabee2010,Lelievre:2012tg,Leimkuhler:2012dp} that achieve some subset of these enumerated desiderata, yet there still is no widespread agreement on an integrator that performs satisfactorily for a broad set of purposes.

\subsection{Resolution: time step rescaling}
Drawing inspiration from several popular integrators, in this paper we derive a simple family of integrators that split the different update types to permit the definition of thermodynamically meaningful quantities for work and heat. With a novel rescaling of the time step, the resulting dynamics preserves five of Pastor \emph{et al.}'s seven dynamical properties for any values of friction and time step (six if fractional time step values are also considered), and furthermore satisfies all of the other desiderata enumerated above, including its utility for nonequilibrium simulations and schemes involving path sampling or reweighting. The resulting integrator \jdc{represents a stochastic generalization of velocity Verlet,} is simple to implement, and could be a general all-purpose replacement for the various discrete-time Langevin integrators now in use.

In sampling contexts, this time step rescaling does slow the exploration of conformational space if the raw integration time step $\dt$ is not concomitantly increased. We believe that such a time step rescaling permits a larger raw time step, but exactly how much larger $\dt$ can be increased (and hence to what extent sampling efficiency can be recovered or even improved) before reaching the stability limit (or some other integration pathology) remains an open question. Addressing this potentially system-specific issue requires further theoretical and numerical investigation, which is beyond the scope of this paper.

\section{Integrator Splitting}
\label{sec:OperatorSplitting}
We derive a family of integrators by splitting the time evolution operator into stochastic and deterministic components~\cite{Bussi2007} and choosing the adjustable parameters to match dynamical quantities from the continuous equations of motion. We write the one-dimensional version here, but the generalization to multiple dimensions is straightforward (see the Supporting Information, ``Multiple dimensions''). The Langevin Liouville operator (sometimes termed the \emph{Liouvillian})~\cite{Zwanzig2001} can be naturally written as a sum of four parts $\mathcal L= {\mathcal L}_{o} + {\mathcal L}_{v} + {\mathcal L}_{\x} +{\mathcal L}_{h}$. The first operator represents stochastic thermalization~\cite{RiskenFP} via an Ornstein-Uehlenbeck operator,
\begin{equation}
{\mathcal L}_{o} = - \gamma \frac{\partial}{\partial v} v -\frac{\gamma}{\beta m} \frac{\partial^2}{\partial v^2} \ ;
\end{equation}
the next two operators represent deterministic Newtonian evolution of velocity and position,
\begin{equation}
{\mathcal L}_{v} = \frac{f}{m} \frac{\partial }{\partial v } \ , \quad 
{\mathcal L}_{\x} = v \frac{\partial }{\partial \x } \ ;
\end{equation}
and the last operator represents the time evolution of the system Hamiltonian according to the predetermined schedule (or \emph{protocol})~$\Lambda$,
\begin{equation}
e^{\dt\, {\mathcal L}_{h} } \hamil(n) = \hamil(n+1) \ ,
\end{equation}
where $n$ denotes the time step index and $t = n \Delta t$ the simulation clock time for time step $\Delta t$. (Note that in the case of a time-independent Hamiltonian, $e^{\dt\, {\mathcal L}_{h} }$ is the identity operator.) Similar to several other integrators,~\cite{Tuckerman1992,Sexton1992,Adjanor2005,Bussi2007,Bussi2009} we approximate the dynamics over a time $\dt$ by applying a series of Strang (symmetric Trotter) operator splittings~\cite{Trotter1959},
\begin{subequations}
\begin{align}
&e^{\dt[A+B+C+D]} \\
&\ = e^{\tfrac{\dt}{2} A} \ e^{\dt\, [B+C+D]} \ e^{\tfrac{\dt}{2} A} + \mathcal{O}(\dt^3) \\
&\ = e^{\tfrac{\dt}{2} A} \ e^{\tfrac{\dt}{2} B} \ e^{\dt\, [C+D]} \ e^{\tfrac{\dt}{2} B} \ e^{\tfrac{\dt}{2} A} + \mathcal{O}(\dt^3) \\
&\ = e^{\tfrac{\dt}{2} A} \, e^{\tfrac{\dt}{2} B} \, e^{\tfrac{\dt}{2} C} \, e^{\dt\, D} \, e^{\tfrac{\dt}{2} C} \, e^{\tfrac{\dt}{2} B} \, e^{\tfrac{\dt}{2} A} + \mathcal{O}(\dt^3) ,
\end{align}
\end{subequations}
where $(A,B,C,D)$ represents a permutation of $({\mathcal L}_{o},{\mathcal L}_{v},{\mathcal L}_{\x},{\mathcal L}_{h})$.  

There are six Strang splittings of the Liouville operators ${\mathcal L}_{o},{\mathcal L}_{\x},{\mathcal L}_{v}$. The Hamiltonian update operator ${\mathcal L}_{h}$ commutes with ${\mathcal L}_{o}$ and with ${\mathcal L}_{\x}$, so for each of these six splittings there are only two unique placements of ${\mathcal L}_{h}$. For each of the six splittings, one placement of ${\mathcal L}_{h}$ interleaves the position, Hamiltonian, and deterministic velocity updates in such a way as to require multiple force evaluations per step, making the scheme computationally inefficient. Thus there are six distinct splittings that each give rise to different finite time step dynamics and require only one force evaluation per step. 
Notably, because the error in each Strang splitting is $\mathcal{O}(\Delta t^3)$, all are identical to the true Liouville operator $\mathcal{L}$ in the limit $\Delta t \rightarrow 0^+$. 

One such splitting is a stochastic generalization of Velocity Verlet that we call OVRVO (to denote the respective ordering of Ornstein-Uehlenbeck (O), deterministic velocity (V), and deterministic position (R) updates (in analogy to the nomenclature of Leimkuhler and Matthews~\cite{Leimkuhler:2012dp}), 
\begin{align}
\label{equ:splitting}
e^{\dt[{\mathcal L}_{o} + {\mathcal L}_{v} + {\mathcal L}_{\x} +{\mathcal L}_{h}]}& \simeq \\
\underbrace{e^{\tfrac{ \dt}{2} {\mathcal L}_{o}}}_a
\underbrace{e^{\tfrac{\dt}{2} {\mathcal L}_{v}}}_b
&\underbrace{e^{\tfrac{\dt}{2} {\mathcal L}_{\x}}}_c 
\underbrace{e^{\dt\, {\mathcal L}_{h} }}_d
\underbrace{e^{\tfrac{\dt}{2} {\mathcal L}_{\x}}}_e 
\underbrace{e^{\tfrac{\dt}{2} {\mathcal L}_{v}}}_f
\underbrace{ e^{\tfrac{\dt}{2} {\mathcal L}_{o}}}_g \ . \notag
\end{align}
The Hamiltonian-update step is placed to minimize the number of force evaluations \jdc{per time step}.  

For this operator splitting, a single update step that advances the simulation clock by $\dt$ is given explicitly by:
\begin{subequations}
\label{equ:OVRVO} 
\begin{align}
v(n+\tfrac{1}{4}) & = \sqrt{a}\ v(n) + \sqrt{\frac{1\text{-}a}{\beta m}} \ \Normal^+(n) \label{equ:heat1} \\
v(n+\tfrac{1}{2})  & = v(n+\tfrac{1}{4}) + \frac{b\, \dt} {2}\  \frac{f(n)}{m} \label{equ:dV1} \\
\x(n+\tfrac{1}{2}) &= \x(n) + \frac{b\,\dt}{2} \, v(n+\tfrac{1}{2}) \label{equ:dX1} \\
\hamil(n) & \rightarrow \hamil(n+1) \label{equ:dH} \\
\x(n+1) &= \x(n+\tfrac{1}{2}) + \frac{b\,\dt}{2}\, v(n+\tfrac{1}{2}) \label{equ:dX2} \\ 
v(n+\tfrac{3}{4}) & = v(n+\tfrac{1}{2}) + \frac{ b\,\dt}{2}\ \frac{f(n+1)}{m} \label{equ:dV2} \\
v(n+1) & = \sqrt{a} \ v(n+\tfrac{3}{4}) + \sqrt{\frac{1\text{-}a}{\beta m}} \ \Normal^-(n+1) \ . \label{equ:heat2}
\end{align}
\end{subequations}
Here, $a=\exp({- \gamma\, \dt})$, and $\Normal^+$ and $\Normal^-$ are independent, normally distributed random variables with zero mean and unit variance (hence, when scaled by $(\beta m)^{-1/2}$, distributed according to the equilibrium Maxwell-Boltzmann velocity distribution). 

The substeps in 
Eq.~\eqref{equ:OVRVO} 
are the finite difference expressions of the corresponding suboperators in 
Eq.~\eqref{equ:splitting}. 
The initial (a) and final (g) operators randomize the velocity and leave the position unchanged, mixing the old velocity with a Maxwell-Boltzmann random variate (with old velocity weighted according to $a = \exp[-\gamma \Delta t]$). These operators can be analytically integrated to give the first 
\eqref{equ:heat1} 
and last 
\eqref{equ:heat2} 
substeps that are stochastic, Markovian, and detailed-balanced (with respect to the true canonical measure\bibnote{Note that, even for a time-independent Hamiltonian, the sampled distribution will differ from the true canonical measure $\mu(r,v) \propto \exp[-\mathcal{H}(r,v)]$ in a time step-dependent manner~\cite{Sivak:2013ke}})~\cite{Uhlenbeck1930, Bussi2007}. The operators (b) and (f) correspond to deterministic velocity updates, while (c)~and~(e) correspond to deterministic position updates. Together they are approximated by the finite difference expressions of substeps [\ref{equ:dV1}, \ref{equ:dX1}, \ref{equ:dX2}, \ref{equ:dV2}], which together constitute the deterministic and symplectic velocity Verlet integrator~\cite{Swope1982, Tuckerman1992}, but slightly altered by an effective time step rescaling $0\leq b\leq 1$, chosen to maintain the continuous-limit zero-force diffusion coefficient and terminal drift under a uniform external force, regardless of friction coefficient or time step (derived in Supporting Information, ``Determination of rescaling parameters'').
In the limit $\dt\rightarrow0$ this rescaling factor $b$ converges to unity and the splitting converges to the continuous-time equation of motion [Eq.~\eqref{equ:contLang}].
Operator (d) and its finite difference expression as 
Eq.~\eqref{equ:dH} 
makes explicit the midpoint Hamiltonian update. Note that several other popular integrators do not explicitly include the update of the system Hamiltonian, presumably because they are not concerned with calculating distributions of the work generated by explicit Hamiltonian changes.

The alternative splittings with one force evaluation per time step include ORVRO (a stochastic generalization of position Verlet~\cite{Tuckerman1992}), 
\begin{align}
e^{\dt[{\mathcal L}_{o} + {\mathcal L}_{v} + {\mathcal L}_{\x} +{\mathcal L}_{h}]}& \simeq \\
e^{\tfrac{ \dt}{2} {\mathcal L}_{o}}
e^{\tfrac{\dt}{2} {\mathcal L}_{\x}}
&e^{\tfrac{\dt}{2} {\mathcal L}_{h}}
e^{\dt\, {\mathcal L}_{v} }
e^{\tfrac{\dt}{2} {\mathcal L}_{h}}
e^{\tfrac{\dt}{2} {\mathcal L}_{\x}}
e^{\tfrac{\dt}{2} {\mathcal L}_{o}} \ ; \notag
\end{align}
RVOVR (an explicit Hamiltonian-update generalization of Leimkuhler and Matthew's `ABOBA'), 
\begin{align}
e^{\dt[{\mathcal L}_{o} + {\mathcal L}_{v} + {\mathcal L}_{\x} +{\mathcal L}_{h}]}& \simeq \\
e^{\tfrac{ \dt}{2} {\mathcal L}_{\x}}
e^{\tfrac{\dt}{2} {\mathcal L}_{h}}
&e^{\tfrac{\dt}{2} {\mathcal L}_{v}}
e^{\dt\, {\mathcal L}_{o} }
e^{\tfrac{\dt}{2} {\mathcal L}_{v}}
e^{\tfrac{\dt}{2} {\mathcal L}_{h}}
e^{\tfrac{\dt}{2} {\mathcal L}_{\x}} \ , \notag
\end{align}
VRORV (an explicit Hamiltonian-update generalization of Leimkuhler and Matthew's `BAOAB'), 
\begin{align}
e^{\dt[{\mathcal L}_{o} + {\mathcal L}_{v} + {\mathcal L}_{\x} +{\mathcal L}_{h}]}& \simeq \\
e^{\tfrac{ \dt}{2} {\mathcal L}_{v}}
e^{\tfrac{\dt}{2} {\mathcal L}_{\x}}
&e^{\tfrac{\dt}{2} {\mathcal L}_{h}}
e^{\dt\, {\mathcal L}_{o} }
e^{\tfrac{\dt}{2} {\mathcal L}_{h}}
e^{\tfrac{\dt}{2} {\mathcal L}_{\x}}
e^{\tfrac{\dt}{2} {\mathcal L}_{v}} \ ; \notag
\end{align}
ROVOR,
\begin{align}
e^{\dt[{\mathcal L}_{o} + {\mathcal L}_{v} + {\mathcal L}_{\x} +{\mathcal L}_{h}]}& \simeq \\
e^{\tfrac{ \dt}{2} {\mathcal L}_{\x}}
e^{\tfrac{\dt}{2} {\mathcal L}_{o}}
&e^{\tfrac{\dt}{2} {\mathcal L}_{h}}
e^{\dt\, {\mathcal L}_{v} }
e^{\tfrac{\dt}{2} {\mathcal L}_{h}}
e^{\tfrac{\dt}{2} {\mathcal L}_{o}}
e^{\tfrac{\dt}{2} {\mathcal L}_{\x}} \ , \notag
\end{align}
and VOROV,
\begin{align}
e^{\dt[{\mathcal L}_{o} + {\mathcal L}_{v} + {\mathcal L}_{\x} +{\mathcal L}_{h}]}& \simeq \\
e^{\tfrac{ \dt}{2} {\mathcal L}_{v}}
e^{\tfrac{\dt}{2} {\mathcal L}_{o}}
&e^{\tfrac{\dt}{2} {\mathcal L}_{\x}}
e^{\dt\, {\mathcal L}_{h} }
e^{\tfrac{\dt}{2} {\mathcal L}_{\x}}
e^{\tfrac{\dt}{2} {\mathcal L}_{o}}
e^{\tfrac{\dt}{2} {\mathcal L}_{v}} \ . \notag
\end{align}

\begin{figure}[h]
\includegraphics{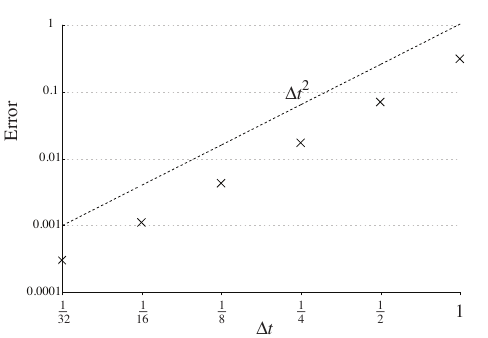}
\caption{{\bf Numerical demonstration that the errors in energy of the OVRVO integrator 
[Eq.~\eqref{equ:OVRVO}] 
are second order in $\dt$.}
Here, we use a previously described model system~\cite{Vanden-Eijnden2006} of a harmonic potential, with unit spring constant, friction coefficient, temperature, and mass, with initial conditions $\x(0) = v(0) = 0$. The error is the absolute deviation of the estimate of $\langle \x^2(1)+v^2(1) \rangle =0.9796111900\ldots$ (twice the energy) computed by ensemble averaging over $10^8$ independent realizations. Standard errors of the mean are substantially smaller than the symbol size. The line is the graph of the function~$\dt^2$.}
\label{order}
\end{figure}

Since for a single time step $\dt$ the error is $O(\dt^3)$ for any of these Strang splittings, when applied over $N = t/\dt$ time steps the global error is $O(\dt^2)$. 
Fig.~\ref{order} 
confirms that OVRVO errors in the energy are second order in the time step~$\dt$. 

\section{Time step rescaling recovers dynamical properties}
\begin{figure}[h]
\includegraphics{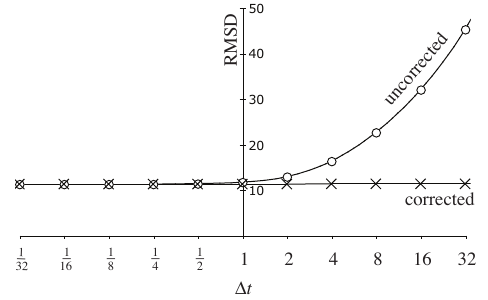}
\caption{{\bf Time step rescaling recovers correct field-free diffusion as a function of time step.}
Root mean-squared displacement versus relative time step length $\dt$ at time $t=64$ for a freely diffusing particle in one dimension, with unit mass, temperature, and friction coefficient, subject to the OVRVO integrator 
[Eq.~\eqref{equ:OVRVO}] 
without time step rescaling, 
$b=1$ ($\circ$), or with time step rescaling,
Eq.~\eqref{tau} 
($\times$). }
\label{drift}
\end{figure}

Standard integrators implicitly set the parameter $b$ to unity in \ref{equ:dV1}, \ref{equ:dX1}, \ref{equ:dX2}, \ref{equ:dV2}. However, we show later that a non-unit $b$ recovers, for arbitrarily-large time step, the continuous-limit values of important dynamical quantities. The time step rescaling can be most simply derived by noting that in the absence of a potential, for any of the six splittings, a trajectory is isomorphic to a semi-flexible Gaussian polymer chain~\cite{Yang2002}: the set of positions correspond to the beads on the polymer chain, and the displacement vectors during single time steps correspond to the inter-monomer bond vectors in the chain. In one dimension, the displacement is a normally distributed random variable with zero mean and variance $\sigma_0^2=(b\dt)^2/(\beta m)$ (in accordance with Maxwell-Boltzmann velocity statistics and the time interval $b\,\dt$), and the autocorrelation between velocities separated by $N$ steps decays exponentially as
\begin{align}
{\big\langle v(N)\ v(0) \big\rangle} = \frac{ a^N}{\beta m} \ .
\end{align}
For this system the mean square displacement in the large time ($\gamma t \gg 1$) limit is~\cite{Yang2002}
\begin{subequations}
\begin{align}
\label{rmsd}
{\big\langle [ \x(N)- \x(0)]^2\big\rangle} &=N\ \sigma_0^2 \ \frac{1+a}{1-a} \ \\ 
& = N \frac{(b\dt)^2}{\beta m} \coth \frac{\gamma \dt}{2} \ .
\end{align}
\end{subequations}
The time step rescaling $b$ results from equating this expression to the mean-squared displacement of a freely diffusing particle in one dimension, $2 D t = 2 t/(\beta m \gamma)$, for a total simulation time over $N$ steps, $t= N\, \dt$. In particular, the time step used in the position update step is rescaled by the factor 
\begin{equation}
b = \sqrt{ \frac{2}{\gamma\,\dt}\tanh\frac{\gamma\,\dt}{2}} \ , \label{tau}
\end{equation}
ensuring that the effective free-particle diffusion constant is independent of time step length (see 
Fig.~\ref{drift}).
In the low friction limit $\gamma\,\dt \ll1$,  $b = 1 - O([\gamma\,\dt]^2)$, and in the high friction limit $\gamma\,\dt \gg1$, $b = \sqrt{2\,/\, (\gamma\dt)}$. 

Note that even though the position update utilizes an effective time step of $b \Delta t$, the simulation clock is still advanced by the full time step $\Delta t$. \jdc{The zero-force MSV and VAC and the uniform-force terminal drift are unaffected by the choice of $b$.} We derive the time step rescaling from a different perspective and in more detail in Supporting Information, ``Determination of rescaling parameters.''

\section{Integrator properties}

\subsection{OVRVO generalizes other popular integrators}
For an explicitly time-independent system Hamiltonian, this family of integrators reduces to various other schemes in certain limits or approximations. At zero friction, $\gamma=0$ and $a=b=1$, thus stochastic substeps have no effect, so OVRVO, VRORV, and VOROV are identical to the deterministic velocity Verlet integrator, whereas ORVRO, RVOVR, and ROVOR are identical to position Verlet. In the high-friction or long-time limit, $a=0$ and $b = \sqrt{2\,/ \, (\gamma \dt)}$, and OVRVO reduces to the Euler integrator for overdamped Langevin dynamics~\cite{Ermak1974} (also known as the Euler-Maruyama method~\cite{Maruyama:1955wn}),
\begin{equation}
\x(n+1) = \x(n) + \frac{\dt}{\gamma} \frac{f(n)}{m} + \sqrt{\frac{2\,\dt}{\beta m\gamma}} {\mathcal N}(n) \ .
\end{equation}
ROVOR and VOROV interpose a velocity randomization substep between the deterministic velocity and position updates. In this $\gamma\dt \gg 1$ limit, the velocity is completely randomized before each position-update substep, and thus the position updates are completely independent of the Hamiltonian. The other four splittings preserve the influence of the Hamiltonian on the dynamics even in this limit of large friction (or large time step). 

OVRVO also reduces to several other popular integrators in other limits or approximations. If the effective time step rescaling for the deterministic substeps is omitted, such that $b = 1$, then OVRVO is equivalent to an integrator described by Adjanor, Ath\`enes, and Calvo~\cite{Adjanor2006}; and by Bussi and Parrinello~\cite{Bussi2007}. If we also combine all stochastic and deterministic velocity updates [\ref{equ:dV2}, \ref{equ:heat2}, \ref{equ:heat1}, \ref{equ:dV1}], we recover the integrator of Ath\`enes~\cite{Athenes2004}; and recasting the Ath\`enes integrator as a Verlet-style integrator (only monitoring position) we converge with the Br\"unger-Brooks-Karplus (BBK) integrator~\cite{Brunger1984} in the low friction limit. If instead we only combine adjacent pairs of stochastic and deterministic velocity updates 
[\eqref{equ:heat1} with \eqref{equ:dV1}, \eqref{equ:dV2} with \eqref{equ:heat2}] 
(still with no time step rescaling) we produce the low friction limit of the Langevin Leapfrog integrator of Izaguirre, Sweet, and Pande~\cite{Izaguirre:2010ve,minh-chodera:jcp:2009:path-reweighting}.

\subsection{Nonequilibrium work}
There is significant interest in probing the probability distribution of work required during a nonequilibrium driving process, which via the work fluctuation theorems~\cite{Jarzynski:2011hl} can report on various system properties. Such usage requires careful splitting of thermodynamically distinct energy changes~\cite{Sivak:2013ke}. 

The total energy change $\Delta E$ during the $n$th full time step of OVRVO can be cleanly separated into heat $\heat$, protocol work $\protwork$, and shadow work $\shadowwork$~\cite{Sivak:2013ke}:
\begin{subequations}
\label{energyDefs}
\begin{align}
\label{firstlaw}
\Delta E   &= \heat+\work \\
& = \heat + \protwork + \shadowwork \\
\heat   &=  \KE \big(v(n+\tfrac{1}{4}) \big) - \KE \big(v(n) \big) \label{defineHeat}\\
&\ \ \ \ \ + \KE \big(v(n+1) \big) - \KE \big(v(n+\tfrac{3}{4}) \big) \notag\\
\protwork  &=   \PE \big(r(n+\tfrac{1}{2}),n+1 \big) - \PE \big(r(n+\tfrac{1}{2}), n\big) \label{defineProtWork} \\
\shadowwork  &=  \KE \big(v(n+\tfrac{1}{2}) \big) - \KE \big(v(n+\tfrac{1}{4}) \big) \label{defineShadowWork}\\
&\ \ \ + \PE \big(r(n+\tfrac{1}{2}), n )\big) - \PE \big(r(n), n\big) \notag\\
&\ \ \ + \PE \big(r(n+1), n+1 \big) - \PE \big(r(n+\tfrac{1}{2}), n+1\big) \notag\\
&\ \ \ + \KE \big(v(n+\tfrac{3}{4}) \big) - \KE \big(v(n+\tfrac{1}{2}) \big) \ . \notag
\end{align}
\end{subequations}
Here, $\PE(\x,n)$ is the potential energy for configuration $\x$ under Hamiltonian $\hamil(n)$, and $\KE(v)=\tfrac{1}{2}m v^2$ is the kinetic energy for velocity $v$. The five other splittings permit similar decompositions of energy changes into heat, protocol work, and shadow work.

\subsection{Constraints} 
Constraints, such as rigid bond lengths, can be readily incorporated into the dynamics using standard techniques~\cite{Lelievre:2012tg}. The symplectic part of the integrator can be constrained with the standard RATTLE [\ref{equ:dV1}, \ref{equ:dV2}] algorithm~\cite{Andersen1983, Frenkel2002}. Since RATTLE is symplectic if iterated to convergence~\cite{Leimkuhler1994}, adding constraints does not interfere with the underlying reversibility of the dynamics. Similarly, the velocity randomization substeps, 
Eqs.~\eqref{equ:heat1} and \eqref{equ:heat2}, 
can be constrained with RATTLE, which modifies the heat flow, but preserves detailed balance~\cite{Lelievre:2012tg}. Consequently, constrained versions of this family of integrators still obeys the precepts of nonequilibrium thermodynamics~\cite{Sivak:2013ke}, with the same definitions of heat, protocol work, and shadow work 
[Eq.~\eqref{energyDefs}], 
provided that the definition of free energy is altered to account for the constrained degrees of freedom~\cite{Lelievre:2012tg}.

\subsection{Computational efficiency}
\label{sec:ComputationalEfficiency}
All six splittings require one force evaluation per time step.
For OVRVO, for example, the force in 
substep~\eqref{equ:dV2} 
is identical to the force in 
substep~\eqref{equ:dV1} 
of the next time interval. Measuring heat requires two evaluations of kinetic energy per time step for all six splittings, for OVRVO just after 
substep~\eqref{equ:heat1} 
and just before 
substep~\eqref{equ:heat2}. 
Separately measuring protocol work and shadow work requires two potential energy evaluations per time step, once each just before and after the Hamiltonian-update substep. Shadow work measurement also requires the kinetic energy evaluations already needed to measure heat, as shadow work and heat are the only processes that change the kinetic energy.

However, if only the total thermodynamic work $\work =\Delta E - Q$ is of interest, this can be calculated given the heat $Q^{(n)}$ during each step $n\rightarrow n+1$
[easily accumulated during integration using 
Eq.~\eqref{defineHeat}]
and the total energy at the beginning and end of the simulation, 
\begin{align}
\notag \work &= \PE\big(r(N),N\big) + \KE\big(v(N)\big) - \PE\big(r(0),0\big) - \KE\big(v(0)\big) \\ & \ \ \ \ \ - \sum_{n=0}^{N-1} Q^{(n)}\ .
\end{align}

The OVRVO integrator requires two normal random numbers per velocity per time step, one each for the initial
[Eq.~\eqref{equ:heat1}]
and final
[Eq.~\eqref{equ:heat2}] 
velocity randomizations. Splitting the velocity randomization across time steps ensures that the dynamics is microscopically reversible and Markovian, and that the induced Markov chain is irreducible. 
(The two separate randomization steps permit the independent adjustment of the velocity and the position to arbitrary values~\cite{Lelievre2010a}.)
ORVRO and VOROV also induce irreducible Markov chains.  RVOVR, VRORV, and ROVOR, by contrast, do not generate irreducible Markov chains: they effectively agglomerate the two velocity randomizations into a single randomization involving one random number, so for a particular new velocity only one position is possible. Irreducibility has utility for path sampling~\cite{Dellago1998,Dellago2002,stoltz:j-comput-phys:2007:path-sampling,Athenes:2008jm} and path reweighting~\cite{minh-chodera:jcp:2009:path-reweighting,chodera:jcp:2011:dynamical-reweighting} schemes, since any proposed discrete time trajectory through phase space is a valid trajectory of an irreducible Markov chain. However, many practical applications do not require \jdc{strict} irreducibility, so one can halve the number of required random variables for OVRVO and ORVRO by combining the last stochastic substep of one full step with the first stochastic substep of the next full step, and for RVOVR and VRORV by combining the two stochastic substeps of a given full step. ROVOR and VOROV separate their stochastic substeps such that they cannot be easily combined. 
Leimkuhler and Matthews show that in the high friction limit and at medium time step VRORV with this single velocity randomization (and time-independent Hamiltonian) is second-order accurate when other integrators become first-order~\cite{Leimkuhler:2012dp}.

When only the total thermodynamic work is of interest, we can combine the last two velocity updates of 
Eq.~\eqref{equ:OVRVO} 
with the first two updates from the next step, and combine the two position updates, to give a three substep stochastic Leapfrog integrator:
\begin{subequations}
\label{equ:OVRVO2} 
\begin{align}
v(n+\tfrac{1}{2}) &= a\, v(n-\tfrac{1}{2})  +\sqrt{\frac{1\text{-}a^2}{\beta m}}\ \Normal(n) \\ 
& \hspace{7em} + (1+a) \frac{b\, \dt} {2}\  \frac{f(n)}{m} \notag \\
\x(n+1) &= \x(n) + b\,\dt\, v(n+\tfrac{1}{2}) \\
\hamil(n) &\rightarrow \hamil(n+1) \ .
\end{align}
\end{subequations}
Under these circumstances, RVOVR, ROVOR, and VOROV reduce to similar three substep integrators, but, due to their sequencing of substeps, ORVRO and VRORV each only reduce to a five substep integrator.

\subsection{Path action}
\label{sec:PathAction}
The path action $\mathcal S[X]$ is a necessary quantity for many path sampling~\cite{Dellago1998,Dellago2002,stoltz:j-comput-phys:2007:path-sampling,Athenes:2008jm} and path reweighting~\cite{minh-chodera:jcp:2009:path-reweighting,chodera:jcp:2011:dynamical-reweighting} techniques. The conditional path probability functional is a product of single time step probabilities,
\begin{subequations}
\label{equ:prodSingleProbs}
\begin{align} 
 \PP{\traj }{x(0),\Lambda } &= e^{-{\mathcal S}[\, \traj \, | \, x(0),\Lambda \, ]} \\ 
&= \prod_{n=0}^{N-1} \PP{ x(n+1) }{ x(n) } \ .
\end{align}
\end{subequations}
Here, $\traj$ is a trajectory through phase space between $x(0) \equiv \{\x(0),v(0) \}$ at time $0$ and $x(N\dt)$ at $N\dt$. Each time step probability is determined by the probability of the requisite random variables, which for OVRVO is
\begin{align}
\label{equ:inRVs}
& \PP{ x(n+1)}{ x(n)} \\
\notag &\ \ \ \ \ \ \ \ \ \ = \frac{\beta m}{b\,\dt(1-a)} \ p\big( \Normal^{+}(n) \big)\ p\big( \Normal^{-}(n+1) \big) \ .
\end{align} 
The first factor $\beta m / [b\,\dt(1-a)]$ is the Jacobian for the change of variables from $\{\x(n+1),v(n+1)\}$ to $\{\Normal^{+}(n),\Normal^{-}(n+1)\}$, and the probabilities are normal with zero mean and unit variance,
\begin{equation}
\label{equ:gaussianRV}
p\big( \Normal^{\pm} \big)= \frac{1}{\sqrt{2\pi}}\exp\left[ -\frac{1}{2} \left({\Normal^{\pm}}\right)^2 \right]\ ,
\end{equation}
where
\begin{subequations}
\label{equ:inMidVs}
\begin{align}
\Normal^{+}(n) &= \sqrt{\frac{\beta m}{1\text{-}a}} \big[ v(n+\tfrac{1}{4})- \sqrt{a}\ v(n) \, \big] \\
\Normal^{-}(n+1) &= \sqrt{\frac{\beta m}{1\text{-}a}} \big[ v(n+1)- \sqrt{a}\ v(n+\tfrac{3}{4}) \big] \ .
\end{align}
\end{subequations}
The intermediate velocities can be determined by the initial and final \jdc{positions and forces}, 
\begin{subequations}
\label{equ:inEndPosVel}
\begin{align}
v(n+\tfrac{1}{4}) &= \frac{1}{b \dt} \big[ \x(n+1) - \x(n) \big]  - \frac{b\dt}{2} \frac{f(n)}{m} \\
v(n+\tfrac{3}{4}) &= \frac{1}{b \dt} \big[  \x(n+1) - \x(n) \big]  +  \frac{b\dt}{2} \frac{f(n+1)}{m} \ .
\end{align}
\end{subequations}
Combining 
Eqs.~(\ref{equ:prodSingleProbs}-\ref{equ:inEndPosVel}) 
gives the action as a function of position and velocity at the unit time steps,
\begin{align}
\label{equ:bigAction}
&\mathcal{S}\big[\, \traj \, | \, x(0),\Lambda \, \big] = \ln \left[\frac{2\pi(1-a)b\dt}{\beta m}\right]^N \\
&+\sum_{n=0}^{N-1} \frac{\beta m}{2(1-a)} \Bigg\{ \left( \frac{\x(n+1)-\x(n)}{b\dt} - \frac{b\dt}{2} \frac{f(n)}{m} - \sqrt{a}\, v(n) \right)^2 \notag \\ 
&+\left( \sqrt{a} \left[\frac{\x(n+1) - \x(n)}{b\dt} + \frac{b\dt}{2} \frac{f(n+1)}{m}\right] - v(n+1) \right)^2 \Bigg\} \ . \notag
\end{align}
The path probability obeys the expected symmetry under time-reversal~\cite{Sivak:2013ke}, where the work is defined as in [\ref{defineProtWork},\ref{defineShadowWork}].

VOROV has a similarly simple expression for the path action. The path action for ORVRO additionally requires the evaluation of the force and its derivative at the half-step position, $\x(n+\half)$, hence requires paths of twice as many points, and therefore is of lesser utility. RVOVR, VRORV, and ROVOR induce \emph{reducible} Markov chains and thus these splittings have infinite path actions for the vast preponderance of paths.

\section{Results}
All of our Strang splittings lead to seven-substep integration schemes that are time-symmetric; are second-order accurate in $\dt$; make Hamiltonian changes explicit; distinguish between heat, protocol work, and shadow work; and easily incorporate constraints. Six of the twelve unique splittings require a single force evaluation per time step and thus are computationally efficient. Setting $a \equiv e^{-\gamma\Delta t}$ and $b \equiv \sqrt{ \frac{2}{\gamma\,\dt}\tanh\frac{\gamma\,\dt}{2}}$ gives for all six one-force-evaluation splittings the continuous-limit MSD, MSV, and VAC in the zero-force case, and the linear-force virial with asymptotic error $O(\dt^2)$. The six splittings differ in the remaining desiderata. We summarize their properties in 
Table~\ref{tab:compChart}.

\setlength{\tabcolsep}{4pt}
\renewcommand\floatpagefraction{.001}
\makeatletter
\setlength\@fpsep{\textheight}
\makeatother
\begin{sidewaystable*}
\centering
\begin{tabular}{ccccccc}
Desideratum & OVRVO & ORVRO & RVOVR & VRORV & VOROV & ROVOR \\
\noalign{\smallskip}\toprule\noalign{\smallskip}
\\
\multicolumn{7}{c}{\em (All six splittings perform identically)} \\
\noalign{\smallskip}\toprule\noalign{\smallskip}
form is time-reversal symmetric & yes & yes & yes & yes & yes & yes \\
splits heat, work, and shadow work & yes & yes & yes & yes & yes & yes \\
easily incorporates constraints & yes & yes & yes & yes & yes & yes \\
force evaluations per time step & one & one & one & one & one & one \\
\noalign{\smallskip}\hline\noalign{\smallskip}
zero-force MSV & exact & exact & exact & exact & exact & exact \\
zero-force VAC & exact & exact & exact & exact & exact & exact \\
zero-force MSD & exact & exact & exact & exact & exact & exact  \\
linear-force virial & $O(\dt^2)$ & $O(\dt^2)$ & $O(\dt^2)$ & $O(\dt^2)$ & $O(\dt^2)$ & $O(\dt^2)$ \\
\noalign{\smallskip}\bottomrule\noalign{\smallskip}
\\
\multicolumn{7}{c}{\em (Splittings differ in performance)} \\
\noalign{\smallskip}\toprule\noalign{\smallskip}
uniform-force terminal drift & exact & exact & exact & exact & $O(\dt^2)$ & $O(\dt^2)$ \\
\multirow{2}{*}{linear-force MSD} & $O(\dt^2)$ at $n$ & $O(\dt^2)$ at $n$ & \multirow{2}{*}{exact at $n$} & \multirow{2}{*}{exact at $n$}  & \multirow{2}{*}{$O(\dt^2)$} & \multirow{2}{*}{$O(\dt^2)$} \\
& exact at $n\!+\!\half$ & exact at $n\!+\!\half$ \\
\multirow{2}{*}{linear-force MSV} & \multirow{2}{*}{exact at $n$} & \multirow{2}{*}{exact at $n$} & $O(\dt^2)$ at $n$ & $O(\dt^2)$ at $n$ & \multirow{2}{*}{$O(\dt^4)$ at $n$} & $O(\dt^2)$ at $n$ \\
& & & exact at $n\!+\!\half$ & exact at $n\!+\!\half$ & & $O(\dt^4)$ at $n\!+\!\half$ \\
\noalign{\smallskip}\hline\noalign{\smallskip}
irreducible Markov chain & yes & yes & no & no & yes & no \\
path action & simple & requires values at $n\!+\!\half$ & may be $\infty$ & may be $\infty$ & simple & may be $\infty$ \\
\noalign{\smallskip}\hline\noalign{\smallskip}
Hamiltonian dependence for large $\gamma\dt$ & yes & yes & yes & yes & no & no \\
can halve number of random variates & yes & yes & yes & yes & no &	 no \\
generalizes several popular integrators & yes & no & no & no & no & no \\
\noalign{\smallskip}\bottomrule\noalign{\smallskip}
\end{tabular}
\caption{Comparison of properties for different splittings. \textnormal{Desiderata are grouped into those satisfied by all six splittings and those where the splittings differ in their performance.}}
\label{tab:compChart}
\end{sidewaystable*}

\section{Conclusions}
As a stochastic generalization of a standard deterministic technique, OVRVO implements what can be considered a form of velocity Verlet with velocity randomization (VVVR). It is an integrator of general utility, satisfying five of Pastor \emph{et al.}'s seven dynamical properties (six if fractional time step quantities are considered), as well as the remaining enumerated desiderata. It is well-suited for the study of nonequilibrium thermodynamics: work is easily measured because the Hamiltonian changes are made explicit, and these measured works are thermodynamically meaningful because OVRVO distinguishes between heat, protocol work, and shadow work. OVRVO's simple form for the path action facilitates its use in trajectory reweighting or path sampling methods. Our novel time step rescaling maintains (for an arbitrary time step) various continuous-limit dynamical quantities, in particular the uniform-force terminal drift and linear-force fluctuations in position and velocity. 
Further research is required to determine to what extent this rescaling permits a larger raw integration time step, and hence how it affects sampling efficiency.
Finally, OVRVO generalizes several popular integration schemes and thus relates naturally to the existing literature. Its extension to a multiple time step integrator~\cite{Schlick:Gl5jUiHQ} is straightforward, requiring only replacement of the inner symplectic step [\ref{equ:dV1}--\ref{equ:dV2}] with a corresponding symplectic multiple time step integrator for deterministic dynamics. 

By contrast, alternative splittings suffer from shortcomings of varying severity and number. ROVOR and VOROV produce uniform-force terminal drift and linear-force fluctuations of position and velocity that differ from continuous-limit values, lose Hamiltonian dependence in the limit of large $\gamma\dt$, and require two random numbers per time step even for reducible Markov chains. RVOVR and VRORV induce Markov chains that are not irreducible and thus these splittings have limited utility for path-sampling schemes. ORVRO seems similarly useful to OVRVO, but in path sampling applications requires paths with twice the number of points as OVRVO.

\acknowledgement
The authors thank Manuel Ath\`enes (Commissariat \`a l'E\'nergie Atomique/Saclay), Gabriel Stoltz (CERMICS, Ecole des Ponts ParisTech), Beno\^{i}t Roux (Univ.\ of Chicago), Jerome P.\ Nilmeier (Lawrence Livermore Natl.\ Lab.), Todd Gingrich (UC Berkeley), Jes\'us A.\ Izaguirre (Univ.\ of Notre Dame), Huafeng Xu and Cristian Predescu (D.\ E.\ Shaw Research), Patrick Varily (Univ.\ of Cambridge), and Michael Shirts (Univ.\ of Virginia) for enlightening discussions and constructive feedback on the manuscript.
J.~D.~C. acknowledges support from a Distinguished Postdoctoral Fellowship from the California Institute for Quantitative Biosciences (QB3) at the University of California, Berkeley. D.~A.~S. and G.~E.~C. were funded by the Office of Basic Energy Sciences of the U.S. Department of Energy under Contract No. DE-AC02-05CH11231. D.~A.~S. was partially supported by NIGMS Systems Biology Center grant P50 GM081879.

\section*{Dedication}
The authors dedicate this work to William C. Swope on the occasion of his 60th birthday and celebration of his seminal contributions to the field of molecular simulation. Additionally, JDC would like to acknowledge the mentorship and insightful advice he has provided over many happy years of collaboration.

%

%


\part*{Supporting Information}

\section{Determination of rescaling parameters}
\label{sec:rescaleParams}
Here we determine the coefficients $a$ and $b$ in the OVRVO equations
[Eq.~7]
by requiring the exact satisfaction of six of the seven dynamical quantities proposed by Pastor, \emph{et al.}~\cite{Pastor:1988dm} (the linear-force virial is not preserved).  Except where noted, the same analysis produces equivalent results for the ORVRO, RVOVR, VRORV, VOROV and ROVOR splittings.

\subsection{Zero-force mean-squared velocity}
The continuous-limit MSV is fixed by the scale of the random velocity fluctuations. In the zero-force case, OVRVO reduces to the following scheme
\begin{subequations}
\begin{align}
v(n+\tfrac{1}{2})  & = a\ v(n-\tfrac{1}{2}) + \sqrt{\f{1\text{-}a^2}{\beta m}}\ \Normal(n) \label{equ:forceFree1}\\
\x(n+1) &= \x(n) + b\,\dt\, v(n+\tfrac{1}{2}) \ . \label{equ:forceFree2}
\end{align}
\end{subequations}
Ensemble averaging after multiplying 
\ref{equ:forceFree1} 
in turn by $\x(n),\x(n+1),v(n+\half)$ produces
\begin{subequations}
\begin{align}
\label{equ:sub4} \langle \x(n) v(n+\half) \rangle &= a\langle \x(n) v(n-\half) \rangle \\
\label{equ:sub5} \langle v^2(n+\half) \rangle &= a\langle v(n-\half) v(n+\half) \rangle + \f{1-a^2}{\beta m} \\
\label{equ:sub6}  \langle v(n-\half) v(n+\half) \rangle &= a\langle v^2(n-\half) \rangle \ . 
\end{align}
\end{subequations}
Angled brackets denote an ensemble average. Combining 
\ref{equ:sub5} and \ref{equ:sub6} 
and rearranging gives:  
\begin{equation}
\langle v^2(n+\half)\rangle - \f{1}{\beta m} = a^2 \l( \langle v^2(n-\half) \rangle - \f{1}{\beta m} \r) \ .
\end{equation}
So for $a<1$ the MSV will asymptotically approach the Maxwell-Boltzmann result $(\beta m)^{-1}$ regardless of its initial value.  

\subsection{Zero-force velocity autocorrelation function}
Requiring the continuous-limit zero-force velocity autocorrelation function fixes the parameter $a$. For discrete Langevin dynamics, averaging after multiplying 
\ref{equ:forceFree1} 
by $v(n+\half-\ell)$ for $\ell > 0$ produces
\begin{equation}
\langle v(n+\half-\ell)v(n+\half)\rangle = a\langle v(n+\half-\ell)v(n-\half) \rangle \ .
\end{equation}
Induction using this and the equilibrium result $\langle v^2(n-\half) \rangle = (\beta m)^{-1}$ reveals that
\begin{equation}
\langle v(n-\half)v(n-\half+\ell)\rangle = \f{a^{\ell}}{\beta m} \ .
\end{equation}
Comparison with the zero-force velocity autocorrelation function for continuous-time Langevin dynamics~\cite{Chandler1987a}
\begin{equation}
\langle v(0)v(t) \rangle = \f{e^{-\gamma t}}{\beta m} \ ,
\end{equation}
fixes $a = e^{-\gamma\Delta t}$. 

\subsection{Zero-force diffusion coefficient \label{sec:TimeScaleCorrection1}}
Requiring the continuous-limit zero-force MSD fixes the time step rescaling $b$ in 
Eqs.~(7c) and (7e).  
Ensemble averaging after multiplying 
\ref{equ:forceFree2} 
in turn by $\x(n),v(n+\half),v(n-\half)$ produces
\begin{subequations}
\begin{align}
\label{equ:sub1} \langle \x(n)\x(n+1)\rangle &= \langle \x^2(n)\rangle + b\Delta t \langle \x(n) v(n+\half) \rangle \\
\label{equ:sub2} \langle \x^2(n+1) \rangle &= \langle \x(n) \x(n+1) \rangle \\
& \ \ \ \ \ \ \ + b\Delta t \langle \x(n+1) v(n+\half)\rangle \notag\\
\label{equ:sub3} \langle \x(n+1) v(n+\half) \rangle &= \langle r(n) v(n+\half) \rangle \\
& \ \ \ \ \ \ \ + b\Delta t \langle v^2(n+\half) \rangle \ . \notag
\end{align}
\end{subequations}
Assuming that the MSV starts (and therefore remains) at 
\begin{equation}
\langle v^2(n+\half)\rangle = \f{1}{\beta m} \ , \ \forall n \ , \label{equ:maxBoltzMSV}
\end{equation}
and that the particle begins at $\x(0)=0$, combining 
\ref{equ:sub3}, \ref{equ:sub4}, and \ref{equ:maxBoltzMSV} 
produces
\begin{subequations}
\begin{align}
\langle \x(n+1) v(n+\half) \rangle &= \f{b\Delta t}{\beta m} \sum_{\bf i=0}^{n} a^i = \f{b\Delta t}{\beta m} \f{1-a^{n+1}}{1-a} \\ 
\langle \x(n) v(n+\half) \rangle &= \f{b\Delta t}{\beta m} \f{a-a^{n+1}}{1-a} \ .
\end{align}
\end{subequations}
Combining these results with 
\ref{equ:sub1} and \ref{equ:sub2} 
gives
\begin{subequations}
\begin{align}
\langle \x^2(n+1) \rangle &= \langle r^2(n) \rangle + \f{(b\Delta t)^2}{\beta m} \l[ 2\f{1-a^{n+1}}{1-a} - 1 \r] \\
&= \f{(b\Delta t)^2}{\beta m} \l[(n+1)\f{1+a}{1-a} - \f{2a(1-a^{n+1})}{(1-a)^2} \r] \\
&\approx \f{(b\Delta t)^2}{\beta m} (n+1)\f{1+a}{1-a} \ , \ n\gg 1 \ .
\end{align}
\end{subequations}
Equating this expression to the (Fickian) MSD in the continuous limit $2 D t = 2 t/(\beta m\gamma)$ of a particle freely diffusing in one dimension during a total simulation time $t= (n+1)\, \dt$, leads to the effective time step rescaling of 
Eq.~(15),
\begin{equation}
b_{\rm Diff} = \sqrt{ \frac{2}{\gamma\,\dt}\tanh\left(\frac{\gamma\,\dt}{2}\right)} \ .
\end{equation}

\subsection{Terminal drift under uniform external force}
\label{sec:TimeScaleCorrection2}
Requiring the continuous-limit terminal drift under a uniform external force fixes $b$ in 
Eqs.~(7b) and (7f). 
The terminal drift equals 
$\langle r(n+1)-r(n)\rangle/\dt = b \langle v(n+\half\dt) \rangle$ 
once transients have died out and this value stops changing. The velocity reaches its terminal value when it (on average) doesn't change over a full step of the integrator, satisfied when
\begin{equation}
\langle v(n-\half) \rangle = \langle v(n+\half) \rangle = a \langle v(n-\half) \rangle + (1+a)\frac{b\dt}{2}\frac{f}{m} \ .
\end{equation}
Thus the terminal drift is 
\begin{equation}
b \ v_{\text term} = \frac{b^2\dt}{2} \frac{f}{m} \coth \l(\f{\gamma\dt}{2}\r) \ .  
\end{equation}
The value of $b$ that equates this to the value for the continuous equations of motion, $f/(m\gamma)$, is
\begin{equation}
\label{equ:bDrift}
b^2_{\rm Drift} = \frac{2}{\gamma\dt} \tanh \l(\f{\gamma\dt}{2}\r) = b^2_{\rm Diff} \ . 
\end{equation}
Thus the time rescaling $b_{\rm Diff}$ restoring the continuous-limit zero-force diffusion constant matches the time rescaling $b_{\rm Drift}$ restoring the continuous-limit terminal drift under a uniform external force, and thus applying a single time rescaling $b = b_{\rm Diff} = b_{\rm Drift}$ maintains the fluctuation-dissipation relation.

The ORVRO, RVOVR, and VRORV integrators also recover the continuous-limit terminal drift $f/(m\gamma)$ for the same $b_{\rm Drift}$, 
\ref{equ:bDrift}. 
VOROV and ROVOR integrators produce a terminal drift that differs from the continuous-limit value by a factor of ${\rm sech}(\gamma\dt/2)$. In particular, in the limit of large $\dt$ the terminal drift goes to zero, and the dynamics are thus insensitive to force. 

\subsection{Mean-squared displacement, mean-squared velocity, and virial for linear force}
For a time-independent harmonic potential $U = \tfrac{1}{2} k \x^2$, the OVRVO integrator reduces to
\begin{subequations}
\begin{align}
v(n+\tfrac{1}{4}) & = \sqrt{a}\ v(n) + \sqrt{\frac{1\text{-}a}{\beta m}} \ \Normal^+(n) \label{equ:heat1Q} \\
v(n+\tfrac{1}{2})  & = v(n+\tfrac{1}{4}) - \frac{kb\, \dt} {2m}\  \x(n) \label{equ:dV1Q} \\
\x(n+\tfrac{1}{2}) &= \x(n) + \frac{b\,\dt}{2} \, v(n+\tfrac{1}{2}) \label{equ:dX1Q} \\
\x(n+1) &= \x(n+\tfrac{1}{2}) + \frac{b\,\dt}{2}\, v(n+\tfrac{1}{2}) \label{equ:dX2Q} \\ 
v(n+\tfrac{3}{4}) & = v(n+\tfrac{1}{2}) - \frac{ kb\,\dt}{2m}\ \x(n+1) \label{equ:dV2Q} \\
v(n+1) & = \sqrt{a} \ v(n+\tfrac{3}{4}) + \sqrt{\frac{1\text{-}a}{\beta m}} \ \Normal^-(n+1) \ . \label{equ:heat2Q}
\end{align}
\end{subequations}
Ensemble averaging after multiplying 
\ref{equ:heat1Q} 
in turn by $v(n+\tfrac{1}{4}), v(n), \x(n)$ produces
\begin{subequations}
\begin{align}
\langle v^2(n+\quar) \rangle &= \sqrt{a} \langle v(n) v(n+\quar) \rangle + \frac{1-a}{\beta m} \\
\langle v(n) v(n+\quar) \rangle &= \sqrt{a} \langle v^2(n) \rangle \\
\langle \x(n) v(n+\quar) \rangle &= \sqrt{a} \langle \x(n) v(n) \rangle
\end{align}
\end{subequations}
This and similar ensemble averages of 
\ref{equ:dV1Q} and \ref{equ:heat2Q} 
produces a system of 18 linear equations for 18 unknowns.  Solving for the MSD and MSV, we find that:
\begin{subequations}
\begin{align}
\langle \x(n+\half)^2 \rangle &= \f{1}{\beta k} \\
\langle \x(n)^2 \rangle &= \f{1}{\beta k} \f{1}{1 - \f{1}{4}\f{k}{m}(b\dt)^2} \\
\langle v(n)^2 \rangle = \langle v(n+\quar)^2 \rangle &= \langle v(n+\tquar)^2 \rangle = \f{1}{\beta m} \\
\langle v(n+\half)^2 \rangle &= \f{1}{\beta m} \f{1}{1 - \f{1}{4}\f{k}{m}(b\dt)^2}
\end{align}
\end{subequations}
So the MSD at half-steps and the MSV at whole-steps match those of the continuous equations of motion.  A similar calculation reveals that ORVRO also produces continuous-limit MSD at half-steps and MSV at whole-steps. Conversely, RVOVR and VRORV produce continuous-limit MSD at whole-steps and MSV at half-steps. 
Researchers have long known of this friction-independent, frequency-dependent rescaling (as compared to the continuous equations of motion) of the position and velocity fluctuations in similar discrete-time integrators,~\cite{Pastor:1988dm} but our proposed remedy of a time step rescaling is (to our knowledge) novel. While this time step rescaling would in principle permit the use of larger integration time steps, in more realistic systems numerical stability may impose sufficiently stringent constraints to preclude a substantially larger time step.

Since for each of these four integrators the MSD and MSV attain their continuous-limit values at different points, the virial at any given point in time always differs from the continuous-limit value for each of these four splittings, with $O(\dt^2)$ error. VOROV and ROVOR integrators produce MSD with $O(\dt^2)$ error, and MSV with $O(\dt^4)$ error at timepoints $n$ and $n+\half$, respectively, and thus virials with at least $O(\dt^2)$ error.

\section{Multiple dimensions}
\label{sec:MultDim}
This family of integrators is trivially extended to multiple degrees of freedom. The position $\x$, velocity $v$, and force $f$ become vectors ${\bf \x}$, ${\bf v}$, and ${\bf f}$, respectively. The mass $m$ becomes a diagonal matrix ${\bf m}$. The normal variates $\Normal^{\pm}(n)$ become random vectors ${\bf N}^{\pm}(n)$, with averages and covariance
\begin{equation}
\l\langle{\bf N}_i^{\rho}(n)\r\rangle = {\bf 0}, \quad \l\langle \Normal_i^{\rho}(m) \Normal_j^{\sigma}(n) \r\rangle = \delta_{ij} \delta_{mn} \delta_{\rho\sigma} \ , 
\end{equation}
where $\rho,\sigma \in \{ +,- \}$. The mass $m$ and coefficients $a$ and $b$ become diagonal matrices ${\bf m}$, {\bf a}, and {\bf b}, respectively, such that $a_{ij}= \delta_{ij} \exp({- \gamma_i\, \dt})$ and
\begin{equation}
b_{ij} = \delta_{ij} \sqrt{ \frac{2}{\gamma_i\,\dt}\tanh\left(\frac{\gamma_i\,\dt}{2}\right)} \ .
\end{equation}
The resulting equations for OVRVO are:
\begin{subequations}
\begin{align}
{\bf v}(n+\tfrac{1}{4}) &= {\bf a}^{1/2} \cdot {\bf v}(n) \\
& \ \ \ + \left( \frac{1}{\beta}({\bf I}\text{-}{\bf a}) \cdot {\bf m}^{-1} \right)^{1/2} \cdot \ {\bf N}^+(n) \notag\\
{\bf v}(n+\tfrac{1}{2})  &= {\bf v}(n+\tfrac{1}{4}) + \frac{\dt}{2} \, {\bf b}\ \cdot {\bf m}^{-1} \cdot {\bf f}(n) \\
{\bf \x}(n+\tfrac{1}{2}) &= {\bf \x}(n) + \frac{\dt}{2} \, {\bf b} \, \cdot {\bf v}(n+\tfrac{1}{2}) \\
\hamil(n) & \rightarrow  \hamil(n+1) \\
{\bf \x}(n+1) &= {\bf \x}(n+\tfrac{1}{2}) + \frac{\dt}{2} \, {\bf b}\, \cdot {\bf v}(n+\tfrac{1}{2}) \\ 
{\bf v}(n+\tfrac{3}{4}) &= {\bf v}(n+\tfrac{1}{2})  + \frac{\dt}{2} \, {\bf b}\ \cdot {\bf m}^{-1} \cdot {\bf f}(n+1) \\
{\bf v}(n+1) &= {\bf a}^{1/2} \cdot {\bf v}(n+\tfrac{3}{4}) \\
& \ \ \ + \left( \frac{1}{\beta}({\bf I}\text{-}{\bf a}) \cdot {\bf m}^{-1} \right)^{1/2} \cdot {\bf N}^-(n+1) \ , \notag
\end{align}
\end{subequations}
for identity matrix ${\bf I}$.

\section{Metropolization and time step scaling}
The shadow work (resulting from the finite time step of discrete Langevin integrators) drives the system out of equilibrium~\cite{Sivak:2013ke}. Metropolization~\cite{Duane1987} is a popular method to recover the equilibrium distribution (though at the cost of unphysical dynamics). For example, in generalized hybrid Monte Carlo~\cite{Horowitz1991, Lelievre2010a}, the shadow work is used in a Metropolis acceptance step, which can improve integrator stability~\cite{BouRabee:2012fk}. In order to maintain a reasonable acceptance rate the time step $\dt$ must scale with the dimension $d$ as $d^{-1/4}$, if the different degrees of freedom are uncorrelated~\cite{Beskos:arxiv}. By similar arguments, when omitting the Metropolis step, achieving statistically robust work-reweighted expectations (or minimizing departures from the desired distribution by keeping the shadow work small in magnitude) also requires scaling the time step $\dt$ as $d^{-1/4}$. As a result, the number of time steps (and hence force evaluations) required to simulate a given time interval to a given accuracy scales as $d^{1/4}$.


\providecommand*\mcitethebibliography{\thebibliography}
\csname @ifundefined\endcsname{endmcitethebibliography}
  {\let\endmcitethebibliography\endthebibliography}{}

\end{document}